\documentclass[%
 aip,
 amsmath,amssymb,
reprint, onecolumn 
]{revtex4-1}

\usepackage{graphicx}
\usepackage{dcolumn}
\usepackage{bm}

\usepackage[utf8]{inputenc}
\usepackage[T1]{fontenc}
\usepackage{mathptmx}
\usepackage{etoolbox}

\makeatletter
\def\@email#1#2{%
 \endgroup
 \patchcmd{\titleblock@produce}
  {\frontmatter@RRAPformat}
  {\frontmatter@RRAPformat{\produce@RRAP{*#1\href{mailto:#2}{#2}}}\frontmatter@RRAPformat}
  {}{}
}%
\makeatother
\begin{document}

\preprint{AIP/123-QED}

\title[]{Numerical Investigation of the Effect of a Magnetic Field on the Transport of Oxygen in Air}
\affiliation{ 
Imperial College London, \\
United Kingdom
}%

\author{A. C. Kruse}
\email{alexander.kruse19@imperial.ac.uk}
\author{P. G. Aleiferis}%
\author{A. Giusti}

\date{\today}

\begin{abstract}
The effects of magnetisation forces in a binary mixture of gases characterised by large differences in magnetic susceptibility are studied using numerical simulations, with a focus on the differential diffusion of the species and the role of the gradient of mixture composition on the flow field resulting from magnetically-induced forces. A quiescent binary mixture of nitrogen and oxygen, representative of air, confined between two parallel plates is considered. In all simulations, a gradient of $\mathbf{B}^2$, the square of the magnetic flux density magnitude, uniform and directed normal to the walls is imposed. Cases characterised by different pressures, different strengths of $\nabla(\mathbf{B}^2)$, and different initial gradients of species composition are investigated, while the same initial temperature is used in all cases. Non-dimensional groups related to the examined configuration are proposed. In cases characterised by an initially uniform mixture composition, species tend to separate and accumulate at opposite walls, due to differential magnetic forces arising from the differences in magnetic susceptibility. For a given strength of $\nabla(\mathbf{B}^2)$, the effect of the magnetic field on the separation of species increases with decreasing pressure. 
In addition to species separation, it is shown that an initial gradient in the mixture composition perpendicular to $\nabla(\mathbf{B}^2)$ induces a significant change in the velocity field, which enhances the transport of species. This effect is due to a lack of alignment between the gradient of averaged magnetic susceptibility and $\nabla(\mathbf{B}^2)$ and could be exploited to achieve targeted mixing using engineered magnetic fields.
\end{abstract}

\maketitle

\section{\label{sec:Intro} Introduction}

The ability to control fluids using electromagnetic fields has long been a goal of engineers and physicists. This led to the development of ferrofluids, where magnetic nanoparticles suspended in a carrier liquid enable precise control over the motion, shape and properties of the fluid through applied magnetic fields. Notable innovations enabled by ferrofluids include advanced loudspeaker technologies, rotary seals, medical drug delivery and microfluidic devices~\cite{KOLE2021168222}, demonstrating the general applicability of electromagnetic control of fluids~\cite{OehlsenFerrofluids2022}. Although ferrofluids have contributed to several revolutionary technologies, the requirement for specialised nanoparticles restricts their use to applications where the introduction of magnetic nanoparticles is acceptable. In addition, their application is limited to the control of liquid fluids, where nanoparticles can be suspended. To achieve magnetic control over a gaseous flow, the magnetic properties of the species in the gas mixture must be exploited. In this context, it is important to note that even the atmosphere surrounding us is composed of a significant percentage of paramagnetic molecules, notably oxygen~\cite{CRCHandbook}, that naturally respond to magnetic fields without necessity of additives. While the magnetic susceptibility of oxygen is orders of magnitude less than that of ferrofluids, various experimental results indicate that a magnetic field can affect the behaviour of oxygen mixtures~\cite{Ueno1990,Gao2024_b,Carruthers1968}. Advances in high field strength electromagnets could further enable forced diffusion of species and direct manipulation of gas mixtures~\cite{Ueno8T1993}. The prospect of controlling gas mixtures and fluids readily available in our environment, such as air, without the addition of nanoparticles represents a paradigm shift that could unlock new technologies in many different applications.


Effects of strong magnetic fields on gaseous flows has been reported in various experiments and exploratory research. The interaction of magnetic fields with non-ferromagnetic fluids has primarily been explored in the context of combustion and electro-chemical applications. The gases of interest are non-conducting, therefore restricting interactions to magnetisation forces acting on magnetic species through magnetic field gradients. In combination with thermal gradients, magnetic field gradients are found to induce significant convection in pure gaseous oxygen~\cite{Carruthers1968}, a phenomenon attributed to the thermal variation of the magnetic susceptibility of oxygen~\cite{JHVanVleck1932}, which is a paramagnetic species. Investigations in isothermal conditions have also shown that magnetic field gradients can alter the shape of gas jets~\cite{Gao2024_b} and cause their stagnation~\cite{Ueno1990}. 
In the context of reacting flows, magnetic forces have been demonstrated to have significant effects on macroscopic properties of flames~\cite{Aoki1989,Khaldi2010,Yamada2003,Kinoshita2004}. A possible application of species diffusion driven by magnetic forces is the local enrichment of oxygen concentration in atmospheric air. Experimental studies have demonstrated the potential of this technology, with up to 0.6\% increase in oxygen concentration being observed~\cite{Cai2008}.

Although numerous studies exist on the effects of magnetic forces on fluid flows, only very few investigations focus on differential diffusion effects induced by magnetic forces in multi-component mixtures composed of species with different magnetic susceptibility. Through a combination of experiments and numerical simulations, Yamada et al.~\cite{Yamada2003} concluded that in the hydrogen-oxygen flame they investigated, the differential diffusion of OH, which is a paramagnetic species present in the reaction zone, is negligible. However, this finding is difficult to generalise without a comprehensive evaluation of the timescales of magnetically driven diffusion. Magnetic field interactions with paramagnetic O$_2$, and the consequent change in the velocity field due to the bulk forces acting on the flow, have been identified as the primary driver of magnetic effects on the majority of combustion applications in the literature~\cite{Yamada2002,Yamada2003}. Note that in these studies, the effect of magnetic fields on differential diffusion of oxygen was not quantified. It should also be noted that, based on the theoretical analysis of magnetisation forces in paramagnetic fluids by Butcher and Coey~\cite{Butcher2022}, a requirement for magnetically induced convection in a wall-bounded flow is that the spatial gradient of mixture averaged magnetic susceptibility and the gradient of the square of the magnetic field magnitude are not aligned. This implies that gradients of mixture composition and their orientation may affect the formation of flow structures, a phenomenon that should be further investigated.

This work investigates the effects of magnetic fields on the transport of species in a canonical configuration comprising an initially stagnating gas mixture between two parallel plates with a magnetic force normal to the plates. The focus is on the differential diffusion of species and the role of mixture gradients on species transport. A binary mixture of N$_2$ and O$_2$ is investigated here, with a composition representative of air. Note that oxygen is the most paramagnetic species present in the air, whereas nitrogen is diamagnetic. 
The objectives of this work are to (i) quantify the effects of magnetically forced differential diffusion on the separation of oxygen from nitrogen for a range of densities of the mixture and strengths of the magnetic field; (ii) identify non-dimensional groups and relevant timescales associated with the diffusion process; (iii) investigate the role of mixture composition gradients on the magnetic force and resulting flow field.
The objectives are addressed through numerical simulations, where the transport equations for species conservation, momentum and energy with magnetic forcing are solved in a computational fluid dynamics framework.

\section{Methods}
In the following, the models used in the present investigation are discussed. The formulation for the magnetic force acting on gaseous species is first presented, followed by the equations solved to study the transport, including differential diffusion, of a mixture of gases with different magnetic properties. Then, the investigated configuration and numerical setup are presented. Non-dimensional numbers are then introduced together with the species equation in non-dimensional form.

\subsection{Magnetic force}
The magnetic force of interest arises due to the magnetisation of different species (e.g., O$_2$, which is paramagnetic). It is assumed that no charged species are present in the mixture, therefore the Lorentz force is zero.
In paramagnetic and diamagnetic materials, which include the gaseous species of interest, and under the assumption of linear material, the magnetisation, $\mathbf{M}$, of the material can be expressed as proportional to the magnetic field $\mathbf{H}$ \cite{Griffiths1989}. %

\begin{equation}
    \mathbf{M} = \chi \mathbf{H},
    \label{eq:Magnetisation}
\end{equation}

\noindent where $\chi$ is a proportionality coefficient called magnetic susceptibility. The magnetic susceptibility is a material property that is dictated by quantum physical phenomena. In the paramagnetic and diamagnetic gases studied in this work, $\chi \sim \mathcal{O}(10^{-6} - 10^{-9})\ll 1$, hence Eq.~(\ref{eq:Magnetisation}) can be approximated as

\begin{equation}
    \mathbf{M} = \frac{\chi}{\mu_0} \mathbf{B},
    \label{eq:approxMagnetisation}
\end{equation}
where $\mu_0$ is the vacuum magnetic permeability. 
The electromagnetic force density applicable to paramagnetic and diamagnetic gases is the Kelvin force, derived through a magnetic dipole model of magnetism~\cite{Butcher2022}, formulated as
\begin{equation}
    \mathbf{F} = \mu_0 (\mathbf{M}\cdot \mathbf{\nabla})\mathbf{H}.
    \label{eq:KelvinForceDensity}
\end{equation}

Combining Eq.~(\ref{eq:KelvinForceDensity}) with Eq.~(\ref{eq:Magnetisation}), and assuming no currents are present in the domain ($\nabla \times \mathbf{H} = 0$ from Maxwell Equations), the force model can be decomposed into two terms:
\begin{equation}
    \mathbf{F} = \nabla \underbrace{\left(\frac{\mu_0}{2} \chi \mathbf{H} \cdot \mathbf{H}\right)}_{P_{mag}} \underbrace{ - \frac{\mu_0}{2} \mathbf{H}\cdot\mathbf{H} \nabla \chi}_{\text{Korteweg-Helmholtz}},
    \label{eq:Korteweg}
\end{equation}

\noindent which is equivalent to Eq.~(\ref{eq:KelvinForceDensity}) under the assumptions outlined above. The first term on the right-hand side of Eq.~(\ref{eq:Korteweg}) is expected to cause a readjustment of the pressure in the fluid flow that is balanced by the walls~\cite{Butcher2022}. The second term is the so-called Korteweg-Helmholtz force, a force which arises due to variations in the concentration of paramagnetic species throughout the domain. In the context of flows with high gradients of concentration of paramagnetic species, this term may become dominant and induce a bulk flow~\cite{Butcher2022}, as will be investigated in this numerical work.

The formulation of the force in Eq.~(\ref{eq:KelvinForceDensity}) can be simplified under the assumption of Eq.~(\ref{eq:approxMagnetisation}), to derive a formulation for the force per unit mass due to a magnetic field gradient:

\begin{equation}
    \mathbf{f_m} = \frac{\chi_\rho}{2 \mu_0} \nabla (\mathbf{B}^2),
    \label{eq:MagneticBodyForce}
\end{equation}
where $\chi_\rho$ is the mass susceptibility (in m$^3$/kg; i.e.~$\chi_\rho = \chi / \rho$, with $\chi$ being the volume susceptibility at the same conditions at which $\rho$ is evaluated). The force formulation in Eq.~(\ref{eq:MagneticBodyForce}) is commonly found in the literature for paramagnetic and diamagnetic species \cite{Yamada2003,Gao2024_b}.  
In the present investigation, experimentally measured values of magnetic susceptibility are used~\cite{CRCHandbook}. For species found in air, the values for $\chi_\rho$ are summarised in Table~\ref{tab:susceptibilities}. Air is approximated as a mixture of O$_2$ and N$_2$, as other species are diamagnetic and have a very low concentration. Note that the literature reports magnetic susceptibilities only for a limited range of species and for a relatively narrow range of temperature. In this study, ambient temperature is used for all investigated cases. The investigation of the effect of temperature on the transport of species under a magnetic field, which includes the modelling of the dependence of $\chi$ on temperature~\cite{Curie1895}, is left for future studies.

\begin{table}[t!]
    \centering
    
    \caption{Magnetic mass susceptibilities, $\chi_\rho$, of common gases found in the atmosphere, in SI units. Magnetic susceptibilities have been taken from Ref.~\cite{CRCHandbook} and converted into mass susceptibilities. The temperature at which the susceptibility of each gas was measured is also reported.}
    \begin{tabular}{ccc} \hline 
         Species of interest& $\chi_\rho$ 
$(\textrm{ SI m$^3$/kg})$& Temperature (K)\\ \hline 
         $\textrm{O}_2$&  $1.355 \times10^{-6}$& 293\\  
         $\textrm{N}_2$&  $-5.38 \times10^{-9}$& 298\\ 
         $\textrm{Ar}$&  $-6.165 \times10^{-9}$& 298\\ 
         $\textrm{CO}_2$&  $-5.999 \times10^{-9}$& 298\\ \hline
    \end{tabular}
    \label{tab:susceptibilities}
\end{table}

\subsection{Mathematical model for the gas mixture}
The investigation is performed by solving the equations for the conservation of mass, species and energy, and the momentum balance for a multicomponent non-reacting gas. These equations take the form proposed by~\cite{PoinsotReacting}:

\begin{itemize}
\item Mass conservation:
\begin{equation}
    \frac{\partial \rho}{\partial t} + \frac{\partial \rho u_i}{\partial x_i} = 0 ;
\end{equation}

\item Momentum balance:
\begin{equation}
    \frac{\partial \rho u_j}{\partial t} + \frac{\partial \rho u_i u_j}{\partial x_i} = -\frac{\partial p}{\partial x_j} + \frac{\partial \tau_{ij}}{\partial x_i} + \rho \sum ^N_{k=1} Y_k f_{k,j};
\end{equation}

\item Conservation of $k$-th species: 
\begin{equation}
\frac{\partial \rho Y_k}{\partial t} + \frac{\partial \rho u_i Y_k}{\partial x_i} = -\frac{\partial \rho V_{k,i} Y_k}{\partial x_i};    
\end{equation}

\item Conservation of total energy (internal and kinetic):
\begin{equation}
    \frac{\partial \rho E}{\partial t} + \frac{\partial \rho u_i E}{\partial x_i} = -\frac{\partial p u_i}{\partial x_i} - \frac{\partial q_i}{\partial x_i} + \frac{\partial \tau _{ij} u_i}{\partial x_j} + \rho \sum^N_{k=1}Y_kf_{k,i}(u_i+V_{k,i}),
\end{equation}
where $q_i$ is given by:
\begin{equation}
    q_i = -\lambda \frac{\partial T}{\partial x_i} + \rho \sum^N_{k=1}h_k Y_k V_{k,i}.
\end{equation}
\end{itemize}
In the equations above, $f_k$ is the force per unit mass acting on species $k$; $N$ is the total number of species in the mixture; for the other quantities the usual notation applies~\cite{PoinsotReacting}. The magnetic force is considered the only volume force acting on the gas and Eq.~(\ref{eq:MagneticBodyForce}) is used to evaluate the magnetic force acting on each species.

The diffusion velocity, $V_k$, in the most general form is found using:
\begin{equation}\label{eq:diff_vel}
    \nabla X_k = \sum ^N _{j = 1} \frac{X_k X_j}{D_{kj}}(V_j - V_k) + (Y_k - X_k) \frac{\nabla p}{p} + \frac{\rho}{p} \sum^N_{j=1}Y_kY_j(f_k - f_j).
\end{equation}
In Eq.~(\ref{eq:diff_vel}), diffusion induced by pressure gradients is often neglected. The term related to body forces is zero if the force per unit mass is the same for all species (e.g., this is the case for the gravitational force). Since there are orders of magnitude of difference between $\chi$ of different gases (e.g., see Table~\ref{tab:susceptibilities}), the term related to volume forces is non-zero in case of magnetic forces. Neglecting the term related to pressure gradient, and using Eq.~(\ref{eq:MagneticBodyForce}) to model the magnetic force per unit mass, the magnetically forced diffusion velocity of species $k$ can be expressed in the following form:

\begin{multline}
    V_k = -\frac{D_k\nabla Y_k}{Y_k} - \frac{D_k\nabla W}{W} + \frac{\rho D_k W_k \nabla(\mathbf{B}^2)}{2 p W \mu_0} \sum^N_{j=1}Y_j(\chi_{\rho,k}-\chi_{\rho,j}) + \\
    \sum^N_{m=1} Y_m\left[\frac{D_m\nabla Y_m}{Y_m} + \frac{D_m \nabla W}{W} - \frac{\rho D_m W_m \nabla (\mathbf{B}^2)}{2 p W \mu_0}\sum^N_{j=1}Y_j(\chi_{\rho,m}-\chi_{\rho,j})\right].
    \label{eq:DiffusionVel}
\end{multline}

These equations were solved using a finite volume approach by means of a custom solver implemented in OpenFOAM-v10~\cite{Openfoam}. The PISO algorithm was used for pressure-velocity coupling. 

\subsection{Configuration and investigated cases}

An initially quiescent mixture of N$_2$ and O$_2$ between two infinite parallel walls separated by a distance $L$ is considered. The effect of magnetic forces on the separation of species and relevant timescales is investigated by considering an initially uniform composition of the gas with a mass fraction of oxygen, $Y_{O2}$, equal to 0.233. To investigate the effect of composition gradients on the formation of flow structures, a linear initial profile of $Y_{O2}$ is imposed (note that this also implies a gradient of nitrogen mass fraction in the opposite direction). In all simulations, a constant and spatially uniform $\nabla (\mathbf{B}^2 )$, oriented perpendicular to the walls in the positive $y$-direction, is applied from $t = 0$ (as if an electromagnet is suddenly switched on after the mixture reaches its initial conditions). This magnetic field induces forces in the direction normal to the walls such that species can accumulate in the vicinity of either wall. Note that the choice of a magnetic field with such characteristics was made to provide a simple reference configuration and is purely hypothetical. The practical implementation of this magnetic field is not necessarily trivial. The feasibility analysis and magnetic circuit design required to achieve such a magnetic field, especially at large spatial scales, are left for future work. Studies have suggested that it is not possible to achieve an exactly constant gradient of the magnetic field in a finite domain~\cite{Quettier2005}. Therefore, the reported results should be considered as an exploratory study to quantify under which magnitudes of $|\nabla(\mathbf{B}^2)|$ diffusion effects are relevant, as no such quantification exists in the literature. 

\begin{table}[t!]
    \centering
\caption{Summary of the investigated conditions. Note that all of the different pressures (densities) were simulated for $|\nabla(\mathbf{B}^2)| = 10^4$ T$^2$/m, whereas a representative case of $p_0 = 10$ kPa was chosen for varying $|\nabla (\mathbf{B}^2)|$. All simulations were performed at $T_0 = 300$ K.}
\label{tab:Numerical Conditions}
\begin{tabular}{lll}
\hline
Quantity & Units & Investigated values \\ \hline
$p_0$ & kPa & $10$; $100$; $1000$ \\
$|\nabla (\mathbf{B}^2)|$ & T$^2$/m & $10^3$; $10^4$; $10^5$; $10^6$ \\ \hline
\end{tabular}
\end{table}

The effect of magnetic forces on the diffusion of species is mainly a function of pressure and temperature of the mixture, as well as of the magnitude of $\nabla(\mathbf{B}^2)$. In this work, the initial temperature, $T_0$, of the mixture is kept the same in all simulations. Therefore, a change in the initial pressure, $p_0$, implies a change in the density of the mixture. As far as investigations with a uniform initial composition of the mixture are concerned, simulations with different values of initial pressure (i.e.~density) and $\nabla(\mathbf{B}^2)$ are performed, as summarised in Table~\ref{tab:Numerical Conditions}, with an initial uniform temperature of 300~K. Pressures were chosen to be representative of a wide range of applications, from air-purification systems to power generation and mobility. The value of $|\nabla(\mathbf{B}^2)|$ quoted in the literature varies significantly according to the field of research. In experiments of flames under magnetic field gradients~\cite{Ueno1990}, the values of $|\nabla(\mathbf{B}^2)|$ typically do not exceed 300 $\textrm{T}^2\textrm{/m}$. However, in the literature on high magnetic separation technologies (related to separating magnetised nanoparticles in fluids), it is reported that values of $|\nabla(\mathbf{B}^2)|$ of $\mathcal{O}(10^6)$ $\textrm{T}^2\textrm{/m}$ are obtainable at small distances~\cite{Ge2017}. As far as simulations with a non-uniform mixture composition are concerned, two cases with a uniform gradient (linear variation) of oxygen composition, parallel and perpendicular to the walls, respectively, are investigated, both at $p_0=10$~kPa, $T_0=300$ K and with $|\nabla(\mathbf{B}^2)|=100$ T$^2$/m. The gradient of mixture composition is initialised such that the volume average composition is equal to the composition of air, and with values of $\partial Y_{O2,0}/\partial x$ or $\partial Y_{O2,0}/\partial y$ (see Figure~\ref{fig:MeshSetup} for a definition of the frame of reference) equal to 4.66.

\subsection{Numerical setup}
\begin{figure}[t!]
\centering
    \includegraphics[width=0.49\linewidth]{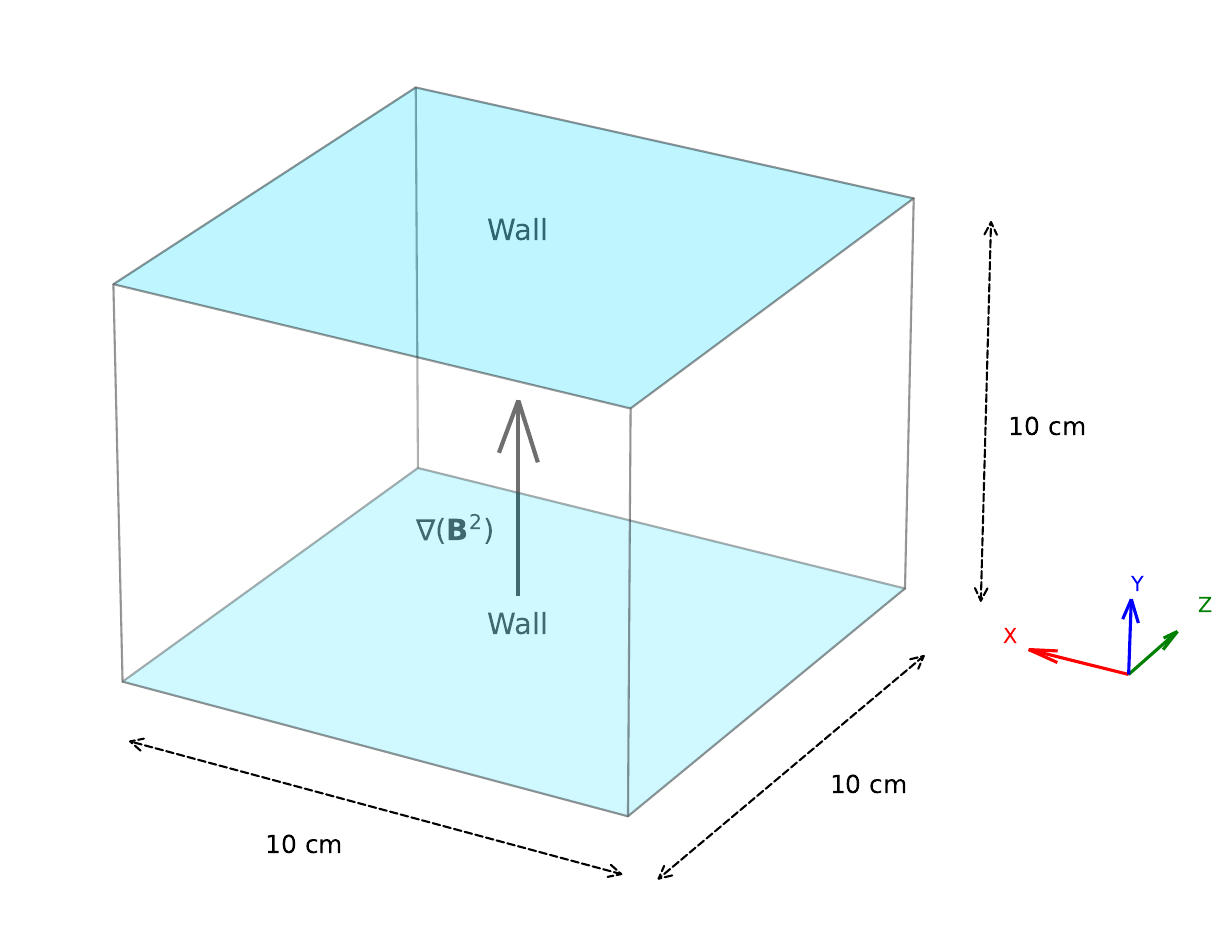}
\hfill
    \includegraphics[width=0.49\linewidth]{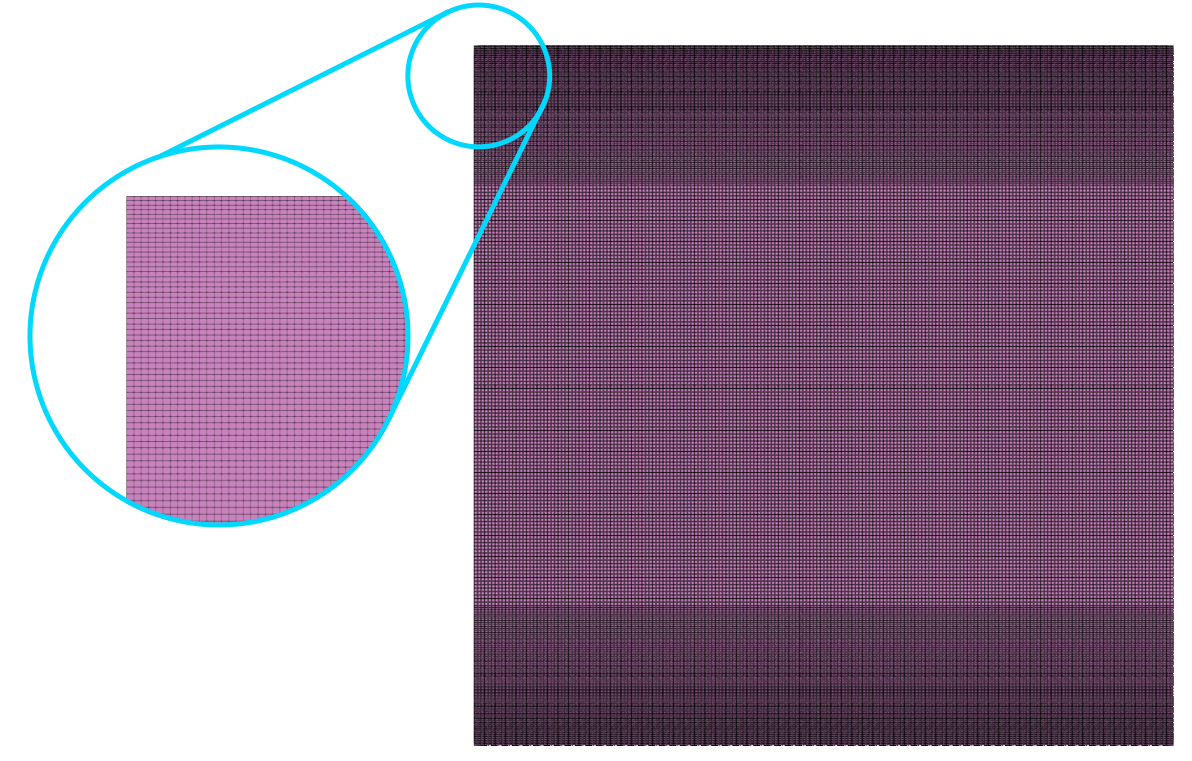}
    
\caption{Schematic of the investigated configuration (left) and computational mesh used in the simulations (right). The inset on the right shows the refinement close to the wall. The frame of reference used in the analysis is also shown; its origin is located at the centre of the domain.}
\label{fig:MeshSetup}
\end{figure}

The configuration is modelled by means of a cuboid of side $L=0.1$~m, with walls on the top and bottom faces, and periodic boundaries on all the other faces. This is schematically shown in Fig.~\ref{fig:MeshSetup}, which also shows the Cartesian frame of reference used for post-processing. The walls are positioned normal to the $y$-direction. The domain is discretised with a hexahedral mesh of 12 million cells. The grid spacing is uniform along $x$ and $z$, while refinements close to the wall are used in the $y$-direction (see inset in Fig.~\ref{fig:MeshSetup}) with a maximum aspect ratio of 3.3 for the first layer of cells next to the wall. Second-order filtered linear schemes were used for species, energy and momentum divergence terms, whereas all the other differential operators were discretised by means of central differencing schemes. A first-order accurate Euler scheme was used for time discretisation. A time step of $10^{-5}$~s was used in all simulations (note that, given the absence of a bulk flow, the CFL number in all simulations was lower than $10^{-3}$).

Custom boundary conditions were implemented for the species and pressure fields at the walls. Specifically, in the case of a non-permeable wall, the boundary condition for the species should ensure that the diffusion velocity normal to the wall is zero. This is achieved by setting the gradient of the $Y_k$ field at the wall so that the component of $V_k$ normal to the wall becomes zero. This condition is obtained by rearranging and simplifying Eq.~(\ref{eq:DiffusionVel}):

\begin{multline}
    \nabla_n Y_k\bigg|_{wall} = \left[- \frac{Y_k\nabla W}{W} + \frac{\rho Y_k W_k \nabla (\mathbf{B}^2)}{2 p W \mu_0} \sum^N_{j=1}Y_j(\chi_{\rho,k}-\chi_{\rho,j}) \right]\cdot  \mathbf{n} ~+ \\
    \frac{Y_k}{D_k}\sum^N_{m=1} Y_m\left[\frac{D_m\nabla Y_m}{Y_m} + \frac{D_m \nabla W}{W} - \frac{\rho D_m W_m \nabla (\mathbf{B}^2)}{2 p W \mu_0}\sum^N_{j=1}Y_j(\chi_{\rho,m}-\chi_{\rho,j})\right]\cdot  \mathbf{n}.
\end{multline}
where $\mathbf{n}$ is the vector normal to the wall boundary and $\nabla_n$ indicates the gradient evaluated along the wall-normal direction. The pressure field also requires an adjusted boundary condition, in order to balance the magnetisation force at the wall boundary. This is implemented as proposed by Ref.~\cite{greenshieldsweller2022}, so that the pressure gradient is equal to the force normal to the boundary:
\begin{equation}
    \nabla_n p \bigg|_{wall} = \rho \mathbf{f_m} \cdot \mathbf{n},
\end{equation}

In order to characterise the evolution of species over time and their potential accumulation close to the wall, the mass fraction of a species of interest is integrated over a series of cut planes parallel to the wall. The integrated quantity for a generic species $k$ and for a plane located at $y=y^*$, defined as
\begin{equation}
    \overline{Y}_k(y^*,t)= \frac{1}{A} \iint_{A} Y_k \, dx\,dz,
\end{equation}
is monitored over time. These planes are distributed equally throughout the direction normal to the wall (i.e. the $y$-direction) to give a spatial profile of the plane-average species mass fraction.

\subsection{Non-dimensional quantities}
In order to identify relevant scales of the problem and generalise the system's behaviour over a wider range of conditions, the following non-dimensional groups can be identified according to the Buckingham-Pi theorem:


\begin{equation*}
    \widetilde{x} = \frac{x}{L}; \qquad  \widetilde{u}_i=\frac{\rho_0 u_i L}{\eta_0}; \qquad  \widetilde{t} = \frac{t \eta_0 }{\rho_0 L^2}; \qquad \widetilde{\rho} = \frac{\rho}{\rho_0};\qquad \widetilde{\nabla}=L \nabla;
\end{equation*}
\begin{equation*}
    \widetilde{D}_k = \frac{\rho_0 D_k}{\eta_0}; \qquad \widetilde{W} = \frac{W}{W_{ref}}; \qquad \widetilde{\chi}_k = \chi_{\rho,k} \rho_0; \qquad \widetilde{p} = \frac{p \rho_0 L^2}{\eta_0^2}; \qquad \widetilde{G} = \frac{\nabla (\mathbf{B}^2) \rho_0 L^3}{\mu_0 \eta_0^2}, 
\end{equation*}

\noindent where $\rho_0$ and $\eta_0$ are the average density and dynamic viscosity of the mixture in the initial state. A further non-dimensional group could be added to represent the initial species gradient in simulations involving a non-uniform initial mass fraction of species. Note also that the investigated configuration is a transport problem with no imposed bulk flow. For configurations involving an imposed bulk flow, different non-dimensional groups can be derived using a characteristic velocity scale. 
Using the proposed non-dimensional quantities, the transport equation for the species mass fraction can be expressed in non-dimensional form as:

\begin{multline}
    \frac{\partial\widetilde{\rho} Y_k}{\partial \widetilde{t}} + \frac{\partial \widetilde{\rho} \widetilde{u}_i Y_k}{\partial \widetilde{x}_i} = \\
\frac{\partial }{\partial \widetilde{x}_i} \widetilde{\rho} Y_k\left\{\frac{\widetilde{D}_k\widetilde{\nabla} Y_k}{Y_k} + \frac{\widetilde{D}_k\widetilde{\nabla} \widetilde{W}}{\widetilde{W}} - \frac{\widetilde{D}_k \widetilde{W}_k \widetilde{\rho} \widetilde{G}}{2 \widetilde{p} \widetilde{W}} \sum^N_{j=1}Y_j(\widetilde{\chi}_k-\widetilde{\chi}_j) ~- \right.
    \left.\sum^N_{m=1}Y_m \widetilde{D}_m\left[\frac{\widetilde{\nabla} Y_m}{Y_m} + \frac{\widetilde{\nabla} \widetilde{W}}{\widetilde{W}} - \frac{ \widetilde{W}_m \widetilde{\rho} \widetilde{G}}{2 \widetilde{p} \widetilde{W}} \sum^N_{j=1}Y_j(\widetilde{\chi}_m-\widetilde{\chi}_j) \right] \right\},
    \label{eq:NonDimensionalSpecies}
\end{multline}
The evolution of species mass fractions as a function of the non-dimensional time is used to provide further understanding of the timescales involved in species diffusion. Verification of the proposed non-dimensional quantities through a scaling of the domain is provided in Appendix~\ref{sec:AppendixScale}. The non-dimensional quantities representing the magnetic field gradient, initial pressure and initial density are summarised in Table~\ref{tab:NonDimnumbers} for all investigated cases.

\begin{table}

\caption{Non-Dimensional quantities corresponding to the cases given in Table~\ref{tab:Numerical Conditions}. Other non-dimensional groups remained identical between different cases. The non-dimensional quantities representing pressure, density and molecular diffusivity of the $k$-th species were evaluated at the initial conditions of the simulations, denoted by a subscript 0; the dynamic viscosity, $\eta_0$, was evaluated for air at $T_0$.
\label{tab:NonDimnumbers}
}
\begin{tabular}{lllp{2cm}p{1cm}p{2cm}p{2cm}p{1cm}}
\hline
Condition Investigated & & \qquad \qquad \qquad \qquad \qquad &$\widetilde{p}_0$ &  $\widetilde{\rho}_0$  & $\widetilde{G}$ & $\widetilde{\chi}_{O2}$ & $\widetilde{D}_{k,0}$ \\ \hline
$p_0 = 10$ kPa& $|\nabla(\mathbf{B}^2)| = 10^4$ T$^2$/m & & 3.402$\times 10^{10}$& 1.0 & 2.702$\times 10^{15}$ &  $1.573\times 10^{-7}$ & 1.0\\
$p_0 = 100$ kPa& $|\nabla(\mathbf{B}^2)| = 10^4$ T$^2$/m & & 3.402$\times 10^{12}$ & 1.0 & 2.702$\times 10^{16}$ & $1.573\times 10^{-6}$ & 1.0\\
$p_0 = 1000$ kPa& $|\nabla(\mathbf{B}^2)| = 10^4$ T$^2$/m & & 3.402$\times 10^{14}$ & 1.0 & 2.702$\times 10^{17}$ & $1.573\times 10^{-5}$ & 1.0\\
$p_0 = 10$ kPa& $|\nabla(\mathbf{B}^2)| = 10^3$ T$^2$/m & & 3.402$\times 10^{10}$ & 1.0 & 2.702$\times 10^{14}$ & $1.573\times 10^{-7}$ & 1.0\\
$p_0 = 10$ kPa& $|\nabla(\mathbf{B}^2)| = 10^5$ T$^2$/m & & 3.402$\times 10^{10}$ & 1.0 & 2.702$\times 10^{16}$ & $1.573\times 10^{-7}$ & 1.0\\
$p_0 = 10$ kPa& $|\nabla(\mathbf{B}^2)| = 10^6$ T$^2$/m & & 3.402$\times 10^{10}$ & 1.0 & 2.702$\times 10^{17}$ & $1.573\times 10^{-7}$ & 1.0\\ \hline
\end{tabular}
\end{table}

\section{Results and Discussion}
\label{sec:results}

\subsection{Simulations with a uniform initial mixture composition}

\begin{figure}[t!]
    \centering
    \includegraphics[width=\linewidth]{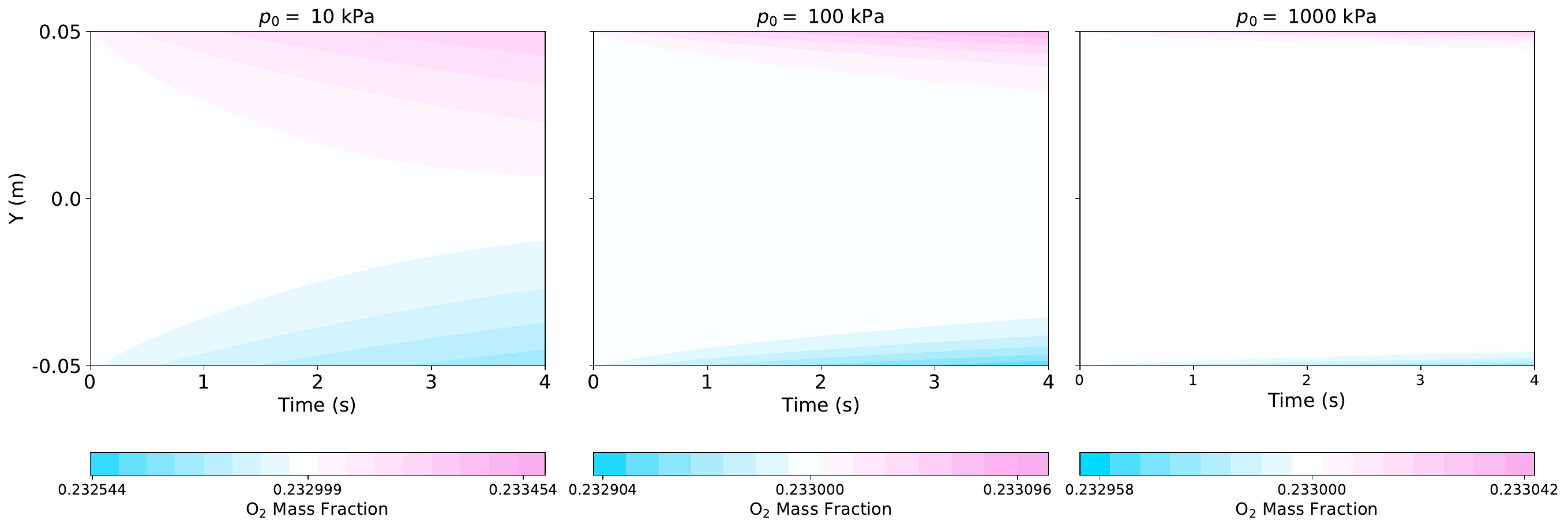}
        \caption{Evolution of oxygen mass fraction (along the $y$-direction) over time for different pressures, at $T_0=300$~K and $|\nabla(\mathbf{B}^2)| = 10^4$ T$^2$/m. Note the different colourbars for each case.}
    \label{fig:colourplot}
\end{figure}

The time evolution of species mass fraction in simulations initialised with a uniform mixture composition is investigated first. Results show that the mass fraction of species tends to be uniform in planes parallel to the wall. Therefore, the time evolution of the mass fraction of one of the mixture components along the $y$-direction gives a good representation of the system behaviour. This is illustrated in Figure~\ref{fig:colourplot}, where the profile of oxygen mass fraction along $y$ as a function of time is shown for cases at $|\nabla(\mathbf{B}^2) |= 10^4$~T$^2$/m and different pressures. Results demonstrate that oxygen tends to accumulate close to the top wall, in the direction of $\nabla(\mathbf{B}^2)$, and to be removed from the bottom wall region. This effect decreases with increasing pressure. Furthermore, the mass fraction distribution is approximately symmetric with respect to the midplane (parallel to the wall) of the domain.

\begin{figure}[t!]
    \centering
    \includegraphics[width=\linewidth]{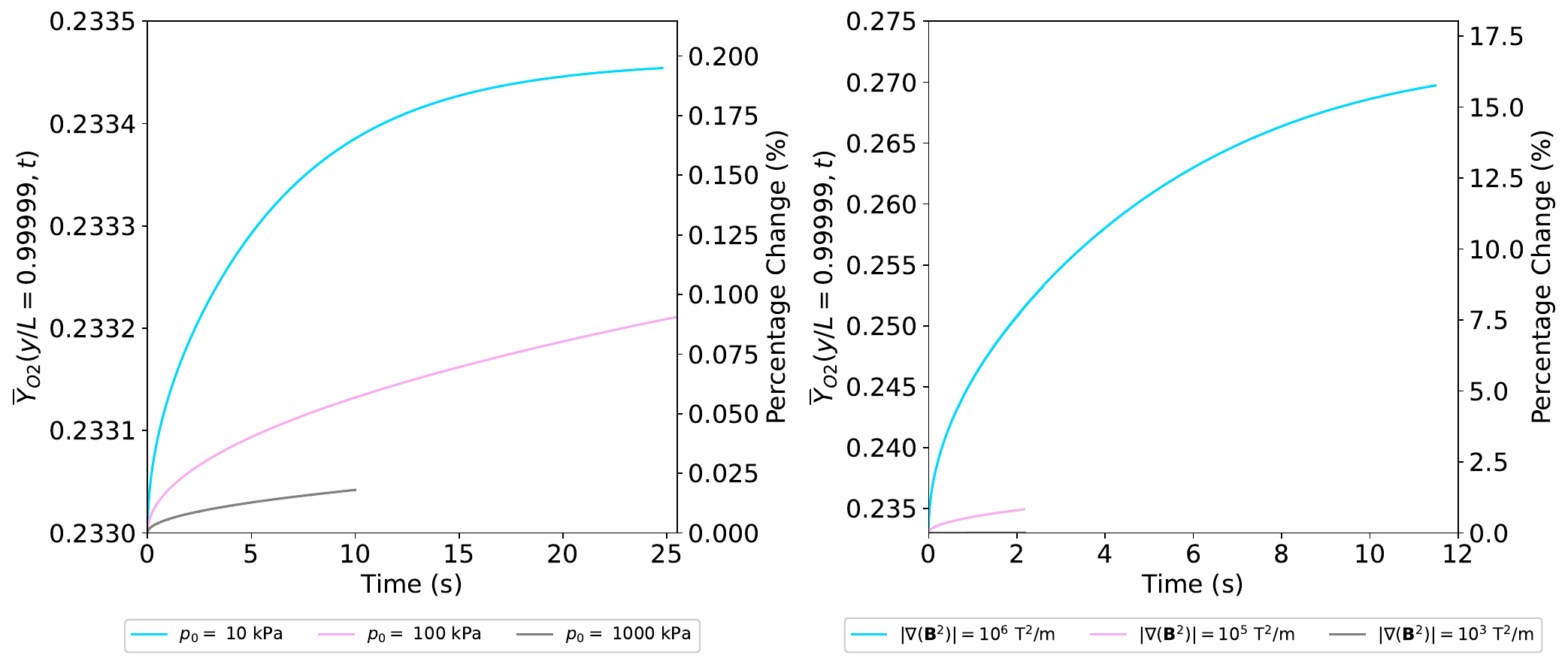}
    \caption{Evolution of oxygen mass fraction over time for different pressures at $|\nabla(\mathbf{B}^2)| = 10^4$ T$^2$/m (left), and for different values of $|\nabla(\mathbf{B}^2)|$ at $p_0 = 10$~kPa (right). The oxygen concentration has been sampled in a plane at $y/L=0.99999$. Note that the black line on the right plot is very close to the time axis.}
    \label{fig:Time_concentration}
\end{figure}

To further compare the time evolutions of composition in the domain, the mass fraction of oxygen in a plane close to the top wall (where the concentration of oxygen is maximum) is shown in Figure~\ref{fig:Time_concentration} for each of the conditions in Table~\ref{tab:Numerical Conditions}. Generally speaking, at lower pressures (i.e., lower densities), a larger separation of the mixture components is observed at the same simulated time. For a given pressure, this separation effect increases with increasing $|\nabla(\mathbf{B}^2)|$ as a consequence of the increase in magnetic forces. For the highest value of $|\nabla(\mathbf{B}^2)|$ investigated here, the oxygen enrichment is higher than 8\% in the simulated time (note that this simulation was performed with an initial pressure of 10~kPa). Such a change in oxygen concentration is quite significant and, for example, may be exploited in applications involving chemical reactions between a fuel and oxygen to change the local oxygen-fuel ratio and therefore the local behaviour of the reacting system. 
It is also interesting to note that the rate of increase in oxygen mass fraction over time decreases with increasing time leading to a `saturation' of the magnetic separation. This is possibly due to a balance between magnetically forced diffusion and molecular diffusion. 
Overall, the results provide evidence that low densities and high magnetic field gradients favour an early separation of oxygen when evaluated with reference to the physical time.

\begin{figure}[t!]
    \centering
    \includegraphics[width=\linewidth]{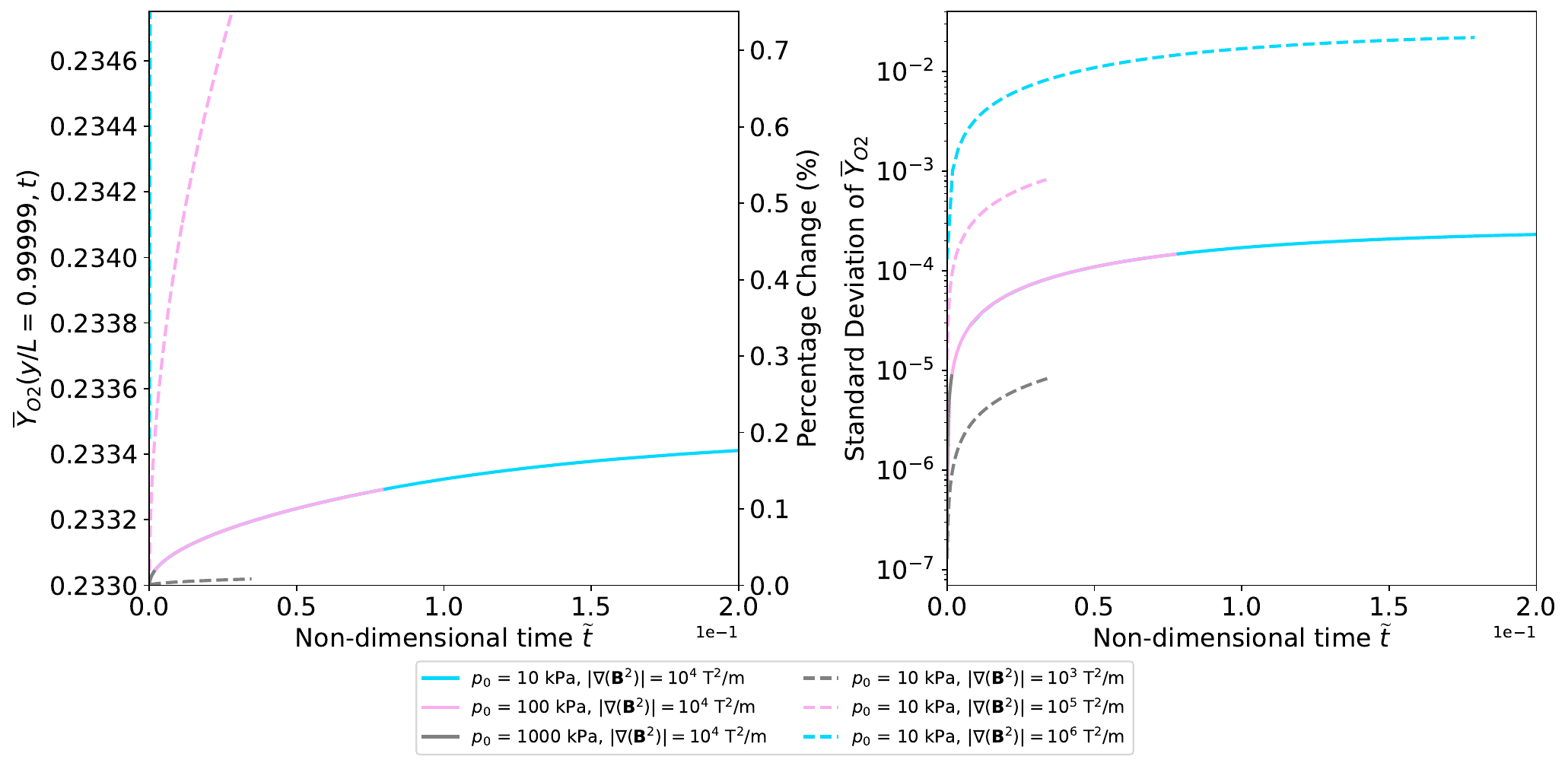}
        \caption{Evolution of the oxygen mass fraction at $y/L=0.99999$ (left) and standard deviation of oxygen mass fraction (evaluated over the $y$-direction, right) as a function of non-dimensional time, for different pressures at $|\nabla(\mathbf{B}^2)| = 10^4$ T$^2$/m, and for various values of $|\nabla(\mathbf{B}^2)|$ at $p_0 = 10$ kPa.}
    \label{fig:deviationEvolution}
\end{figure}



To further analyse the evolution of oxygen separation at different conditions, Figure~\ref{fig:Time_concentration} is made non-dimensional using the derived non-dimensional groups. The evolution of $\overline{Y}_{O2}$ as a function of non-dimensional time in a plane close to the top wall is shown in Figure~\ref{fig:deviationEvolution}, together with the standard deviation of $\overline{Y}_{O2}$ in the domain. The latter provides an alternative metric for evaluating the extent of oxygen separation throughout the domain. As shown in Figure~\ref{fig:deviationEvolution}, cases with the same $|\nabla(\mathbf{B}^2)|$ but different initial pressure (it is reminded that the initial temperature is the same for all cases) are characterised by the same evolution of oxygen accumulation over non-dimensional time. It is of interest to note that all these cases have different values of $\widetilde{G}$ and of the non-dimensional groups representing the initial pressure and mass susceptibility of species (see Table~\ref{tab:NonDimnumbers}); however, the term $\widetilde{G}\widetilde{\chi}/\widetilde{p}_0$ remains the same for all these cases. Inspection of the non-dimensional transport equation for species mass fraction, i.e., Eq.~(\ref{eq:NonDimensionalSpecies}), shows that the term related to magnetic forces is proportional to $\widetilde{G}\widetilde{\chi}/\widetilde{p}$, which justifies the same evolution of oxygen mass fraction in non-dimensional space for all simulations at constant $|\nabla(\mathbf{B}^2)|$ and varying pressure. However, varying $|\nabla(\mathbf{B}^2)|$ at constant pressure and temperature significantly modifies the non-dimensional behaviour of the system, with oxygen separation (at the same non-dimensional time) that increases with increasing magnetic force. 


\begin{figure}[t!]
    \centering
    \includegraphics[width=0.5\linewidth]{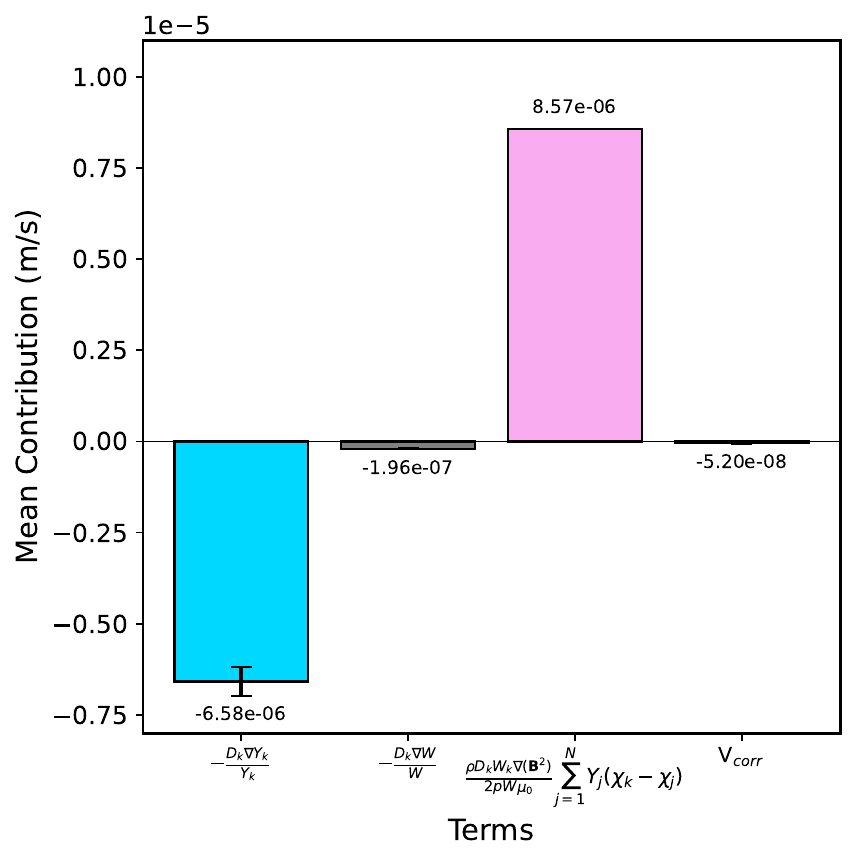}
    \caption{Mean contributions of individual terms of the diffusion velocity of O$_2$, i.e.~Eq.~(\ref{eq:DiffusionVel}) applied to oxygen, sampled along the $x$-centreline of the computational domain, for the case at $T_0 = 300$~K, $p_0 = 10$~kPa and $|\nabla(\mathbf{B}^2)| = 10^4$ T$^2$/m. The analysis has been performed at $t = 24.7$~s, at which the system has reached a quasi-steady state behaviour. V$_{corr}$ refers to the last term in Eq.~(\ref{eq:DiffusionVel}), which is the correction velocity used to ensure mass conservation. The diffusion velocity is obtained by summing the four terms together.}
    \label{fig:TermCostAnalysis}
\end{figure}

To determine the underlying contributions to the forced diffusion velocity of oxygen, the values of the individual terms of Eq.~(\ref{eq:DiffusionVel}) are compared in Figure~\ref{fig:TermCostAnalysis} at a selected time instant. All individual contributions were sampled along the $x$-centreline of the domain. Note that the diffusion velocity is only induced along the $y$-direction. The analysis of the individual terms of the diffusion velocity clearly indicates that the separation of species is driven by the competition between magnetically forced diffusion and molecular diffusion. The two opposing effects result in a diffusion velocity of the order $\mathcal{O}(10^{-6})$ m/s in the $y$-direction (aligned with $\nabla(\mathbf{B}^2)$). The other terms in Eq.~(\ref{eq:DiffusionVel}) are several orders of magnitude smaller. The latter suggests that the spatial variation of the mixture averaged molecular weight has little effect in the investigated cases. The competing effect between magnetically forced diffusion and molecular diffusion is further supported by the fact that if the magnetically forced stratification is let to decay through the removal of the magnetic field, the oxygen near-wall mass fraction will return to the equilibrium in a typical diffusion-like decay (not shown here). 

\subsection{Simulations with a linear variation in the initial mixture composition}

\begin{figure}[t!]
    \centering
    \includegraphics[width=0.49\linewidth]{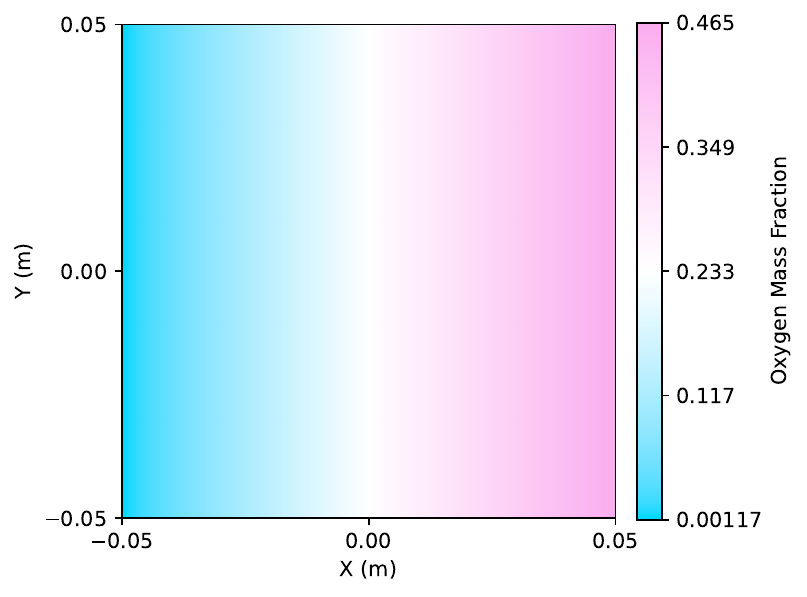} 
    \hfill
    \includegraphics[width=0.49\linewidth]{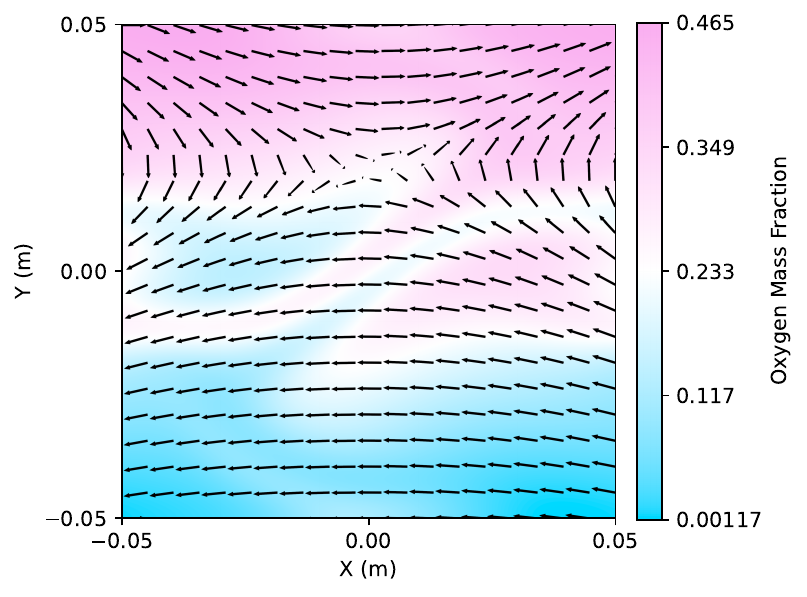}
    \caption{Contour plots of the instantaneous oxygen mass fraction in a plane perpendicular to the walls centred at $z = 0$ for a simulation run with an initial gradient of oxygen mass fraction along the $x$-direction. The left plot shows the $Y_{O2}$ distribution at $t = 0$~s; the plot on the right shows the $Y_{O2}$ distribution at $t = 0.3$~s. Arrows aligned with the induced velocity field are shown on the plot on the right. The investigated condition is at $p_0 = 10$~kPa, $T_0 = 300$~K and $|\nabla(\mathbf{B}^2)| = 10^2$ T$^2$/m.}
    \label{fig:O2Profiles}
\end{figure}

In all cases discussed in the previous section, the body force acting on the fluid is counterbalanced by a pressure gradient, such that no bulk flow is induced by the magnetic forces. This is consistent with the decomposition of the body force into the two components given by Eq.~(\ref{eq:Korteweg}). In the cases characterised by a uniform initial mixture composition, the Korteweg-Helmholtz force (second term on the right-hand side of Eq.~(\ref{eq:Korteweg})) is zero throughout the simulation, whereas the magnetic pressure is balanced by the pressure field. To further study the effect of the Korteweg-Helmholtz force on the flow field, cases with an initial gradient of oxygen mass fraction are investigated next. When simulations are initialised with a gradient of oxygen mass fraction, which implies a non-uniform mixture-averaged $\chi$, the Korteweg-Helmholtz force becomes non-zero from the start of the simulation. However, if the gradient of $\chi$ is aligned with the direction of $\nabla(\mathbf{B}^2)$, the Korteweg-Helmholtz force is counterbalanced by a pressure gradient (as also observed at later stages in simulations initialised with a uniform mixture) and does not induce any bulk flow (this corresponds to the simulation initialised with $\partial Y_{O2,0}/\partial y\neq 0$, not shown here). Conversely, when $\nabla (\mathbf{B}^2) \times \nabla \chi \neq 0$, flow structures are induced by Korteweg-Helmholtz force, as shown in Figure~\ref{fig:O2Profiles}.
In other words, cases where $\nabla(\mathbf{B}^2) \parallel \nabla\chi$ do not show an induced flow, while significant flow structures are observed when $\nabla(\mathbf{B}^2) \perp \nabla\chi$. It is noteworthy that the values of $|\nabla(\mathbf{B}^2)|$ required to induce flow structures in the presence of a gradient in mixture composition are much lower than those required to observe a significant differential diffusion of species within the simulated time. These results support that a non-zero $\nabla \chi$ can induce flow structures in a gaseous mixture when $\nabla \chi$ is not aligned with $\nabla(\mathbf{B}^2)$. It is also worth noting that in the configuration initialised with a gradient in mixture composition perpendicular to $\nabla(\mathbf{B}^2)$, the flow structures evolve in time until the species composition reaches the equilibrium state characterised by a species composition gradient normal to the wall (i.e., in the same direction as $\nabla(\mathbf{B}^2)$).

\section{Conclusions}
The effects of the magnetisation force on the transport of oxygen in air have been investigated using numerical simulations, with a focus on differential diffusion of species and the role of mixture composition gradients in generating flow structures. A reference configuration consisting of air between two parallel plates, initially at rest, has been considered. It is demonstrated that in the case of a uniform $\nabla(\mathbf{B}^2)$ perpendicular to the walls and with a uniform initial composition of the mixture, the difference in magnetic susceptibilities of the species can be used to separate oxygen from nitrogen, leading to a stratification of the species close to the walls. The effect is more prominent at lower densities of the mixture and for higher magnitudes of $\nabla(\mathbf{B}^2)$. Non-dimensional groups are proposed to demonstrate the scaling of the various physical quantities involved in the problem and to generalise the results on species stratification to conditions different from those investigated here. Analysis of the various contributions to the diffusion velocity of species indicates that the stratification effect is primarily caused by a competition between molecular diffusion and magnetically forced diffusion. It is also observed that magnetisation forces may induce flow structures when a gradient in magnetic susceptibility not aligned with $\nabla(\mathbf{B}^2)$ is present. This provides further corroboration of the conditions needed for magnetic forces to induce convection in a gas flow. The present investigation provides fundamental understanding of the effect of magnetic fields on the transport of species in multi-component gases, which has the potential of opening up avenues for new technologies for the control of mixing in gaseous flows, with potential applications reaching high-efficiency oxygen production and medical applications.

\begin{acknowledgments}
We acknowledge the computational resources and support provided by the Imperial College Research Computing Service (doi.org/10.14469/hpc/2232).

\end{acknowledgments}

\section*{Data Availability Statement}

Data available on request from the authors.

\appendix

\section{Simulations with variable $L$}
\label{sec:AppendixScale}

\begin{figure}[t!]
    \centering
    \includegraphics[width=0.7\linewidth]{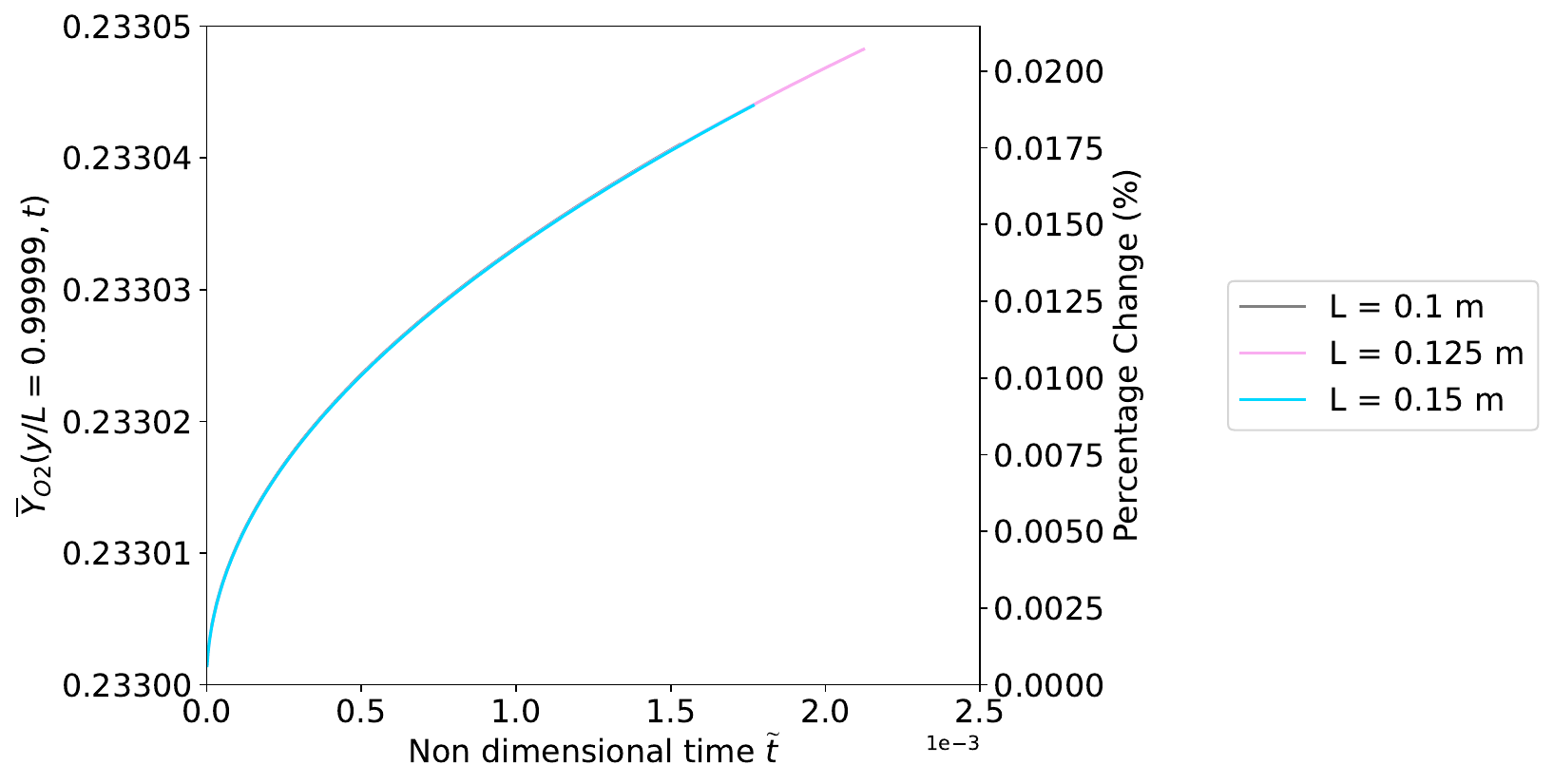}
    \caption{Non-dimensional mass fraction of oxygen for different physical domain sizes, $L\in \{0.1, 0.125, 0.15\}$ m, pressures $p_0 \in \{10^4, 80 \times 10^3, 66.66 \times 10^3\}$ kPa, gradients $|\nabla(\mathbf{B}^2)| \in \{10^4, 6.4\times 10^3, 4.44\times 10^3\}$ T$^2$/m and oxygen mass susceptibilities $\chi_\rho \in \{1.355\times10^{-6}, 1.694\times 10^{-6}, 2.0325\times 10^{-6}\}$ m$^3$/kg respectively. Results have been sampled at a plane $y/L=0.99999$.}
    \label{fig:Non-dimensional}
\end{figure}

In all cases discussed in Section~\ref{sec:results}, the domain length, $L$, is kept the same. To further verify the proposed non-dimensional groups, simulations of three physically scaled cubic domains of side lengths $L = $ 0.1, 0.125 and 0.15 m, respectively, are performed. The base case is the configuration at $p_0 = 10$ kPa, $T_0 = 300$ K, $|\nabla(\mathbf{B}^2)| = 10^4$ T$^2$/m and $L=0.1$~m. Keeping the temperature constant, the values of initial pressure, $\nabla(\mathbf{B}^2)$, and $\chi_\rho$ for each value of $L$ are calculated such that $\widetilde{G}\widetilde{\chi}/\widetilde{p}_0$ are the same as in the base case. The resulting non-dimensional evolutions of the oxygen mass fraction at $y/L= 0.99999$ are shown in Figure~\ref{fig:Non-dimensional}. As expected, these results are identical for the three simulated cases, confirming the proposed non-dimensional scaling. It is interesting to note that for larger domains and the same initial mixture conditions, lower values of $|\nabla(\mathbf{B}^2)|$ are required to obtain the same separation effect, indicating that a given magnetic field is more effective when used in large domains. This has important implications for practical implementation. This is also evident from Eq.~(\ref{eq:NonDimensionalSpecies}) and the term $\widetilde{G}\widetilde{\chi}/\widetilde{p}$. To keep this term constant (same non-dimensional response of the system), an increase in $L$ should be counterbalanced by decreasing $|\nabla(\mathbf{B}^2)|$.

\section*{References}

\bibliography{references.bib}

\end{document}